\documentstyle[11pt,aaspp4]{article}
\def\xxx{\enspace\enspace\enspace}
\def\xxxx{\enspace\enspace\enspace\enspace}

\def\et{et al.}

\def\ha{H$\alpha$}

\def\rhalf{R$_{0.5}$}

\def\solar{\ifmmode_{\mathord\odot}\;\else$_{\mathord\odot}\;$\fi}

\def\HII{H$\,${\sc ii}}

\begin{document}

\title{The Star Clusters in the Starburst Irregular Galaxy 
NGC 1569\footnote{\rm Based
on observations with
the NASA/ESA  Hubble Space Telescope, obtained at the Space Telescope
Science Institute, which is  operated  by AURA, Inc.,
under NASA contract NAS 5-26555.}
}

\author{Deidre A.\ Hunter}
\affil{Lowell Observatory, 1400 West Mars Hill Road, Flagstaff, Arizona 86001
USA;
\\dah@lowell.edu}

\author{Robert W.\ O'Connell}
\affil{University of Virginia, Department of Astronomy, PO Box 3818, 
Charlottesville,
Virginia 22903-0818 USA;
\\rwo@virginia.edu}

\author{J.\ S.\ Gallagher}
\affil{Washburn Observatory, University of Wisconsin, 475 N.\ Charter St.,
Madison, Wisconsin 53706 USA; 
\\jsg@astro.wisc.edu}

\and

\author{Tammy A.\ Smecker-Hane}
\affil{University of California, Department of Physics and Astronomy, 
4129 Reines Hall, Irvine, California 92697-4575 USA; 
\\smecker@carina.ps.uci.edu}

\begin{abstract}

We examine star clusters in the irregular,
starburst galaxy NGC 1569 from {\it HST} images taken with filters
F336W, F555W, and F814W.
In addition to the two super star clusters that
are well known, we identify 45 other clusters that are compact but resolved.
Integrated UVI colors of the clusters span a large range, and comparison with
coeval evolutionary models suggest that the ages range from 2--3 Myrs to
1 Gyr. Most of the clusters have colors consistent with ages of
$\leq$30 Myrs placing them at the tail end
of the recent burst of star formation.

We examine the radial surface brightness profiles of four of the clusters,
and fit King models to three of them.
The colors of the clusters are approximately constant with radius.
The four
clusters have half-light radii and core radii that are in the range observed
in present-day 
globular clusters in our Galaxy. However, 
they are somewhat less concentrated that the average
globular. The two well-known super star clusters have luminosities, 
and one has a known mass,
that are comparable to those of typical globular clusters. The other
two clusters, and likely numerous others in the sample, are similar
to a small globular cluster and to R136 in the LMC.
The conditions that produced the recent starburst, therefore, have
also been those necessary for producing compact, bright star clusters.

We examine resolved stars in the outer parts of the super star clusters.
We find that cluster A contains many
bright blue stars. Some of the blue stars are bright enough to be
evolved massive stars. There is also a small population of red supergiants.
Components A1 and A2 within cluster A have similar colors and
a two-dimensional color map does not offer evidence that one component 
is dominated by red supergiants and the other not. 
The contradiction of the presence of
red supergiants with Wolf-Rayet stars may instead
not be a contradiction at all since there coexistence in a coeval
population is not inconsistent with the evolution of massive stars.
Alternatively, there may be a small age spread 
of several Myrs within cluster A.
The stars that we resolve around cluster B, on the other hand,
contain a small population of more normal blue massive stars and
a large population of red supergiants. The presence of the red
supergiants is consistent with the view that cluster B is in
its red supergiant phase. The presence of the red supergiant stars
in clusters A and B is also verified in near-infrared spectra
where we find strong stellar CO absorption features.
The various age indicators are consistent with a picture in which cluster B
is of order 10--20 Myrs old, 
and cluster A is $\geq$4--5 Myrs old. 
The timescale to form the holes seen in \ha\ and 
HI is comparable to the age of
cluster B. 

\end{abstract}


\section{Introduction}

With the {\it Hubble Space Telescope} ({\it HST}) 
people are finding increasing numbers of super star clusters
(see, for example, 
Holtzman \et\ 1992; 
Whitmore \et\ 1993; 
Conti \& Vacca 1994;
Hunter \et\ 1994; 
O'Connell \et\ 1994, 1995;
Barth \et\ 1995).
These are compact, luminous star clusters that have sizes and
luminosities (when scaled to a common age) that make them comparable
to globular clusters, the most massive star clusters known.
However, unlike globular clusters, the super star clusters being
found today are often young, in some cases as young as a few Myrs.
Therefore, these super star clusters are invaluable in providing us
with a unique window on the early stages, evolution, and conditions
necessary to form globular-type clusters. They also probe the star
formation process at one of its extremes.

Young super star clusters as massive as globular clusters
are not common in
normal disk galaxies,
but there are several of these clusters
within a few Mpc of the Milky Way. Because of their proximity, these clusters
are the best examples for investigating the details of the clusters
themselves. The closest example of a young massive, compact star cluster
is R136, at the
center of the 30 Doradus nebula in the Large Magellanic Cloud (LMC).
Within a cluster radius of 4.7 pc, R136 contains 300 times the concentration
of luminous stars in a typical OB association (O'Connell \et\ 1994, 
Hunter 1995).
R136 is unique among the
globular-like clusters in that it can be resolved into individual
stars and one can investigate the mass function resulting from such a
concentrated star-forming event.
Nevertheless, R136 is 
several magnitudes fainter than other super star
clusters when normalized to the same age. R136 has an integrated
V-band magnitude of only $-$11.
 
Beyond the LMC, the next closest known super star clusters are among the most
extreme in terms of luminosity. These are clusters A and B
located in the nearby peculiar irregular galaxy NGC 1569
(Ables 1968).
These clusters are so compact that they appear stellar
in ground-based images with good seeing,
and their nature was controversial (Arp \& Sandage 1985).
However, {\it HST} Cycle 1 images resolved these objects, proving them to be
star clusters within NGC 1569 with absolute V-band magnitudes of
$-$14 and $-$13 (O'Connell \et\ 1994). 

These super star clusters are resident in the central region of a galaxy
that itself is very unusual.
NGC 1569 
has recently undergone a true wide-scale burst of star formation
involving most of the optical galaxy
(de Vaucouleurs, de Vaucouleurs, \& Pence 1974; Hodge 1974;
Gallagher, Hunter, \& Tutukov 1984; Israel 1988; 
Vallenari \& Bomans 1996; Greggio \et\ 1998). 
The galaxy still today has substantial on-going star formation
as seen by the presence of bright \HII\ regions.
In addition Della Ceca \et\ (1996) have detected hard X-rays 
from supernova remnants and binaries associated with the center of the galaxy
and soft, diffuse X-rays extending along the minor axis of the galaxy
(see also Heckman \et\ 1995), both consequences of the intense recent
star formation. Furthermore, there
are large filaments of ionized gas visible in \ha\ that extend to
1.9 kpc from the center of the galaxy 
(Waller 1991; Hunter, Hawley, \& Gallagher 1993;
Hunter \& Gallagher 1997) and a complex velocity field in the ionized
gas (Tomita, Ohta, \& Sait\_o 1994).
Stil \& Israel (1998) have detected a $7\times10^6$ M\solar\ HI cloud
at 5 kpc from NGC 1569 and the hint of a bridge in HI connecting them.
The suggestion is that interaction with this HI cloud is responsible for the
recent high level of star formation.
Furthermore, it is perhaps a consequence of these unusual circumstances
that this galaxy has made two of the rare super star clusters.

In order to examine the structure of these super star clusters in NGC 1569,
we obtained post-refurbishment images of the galaxy with {\it HST}.
With these data we intended to test the suggestion that the
super star clusters in NGC 1569 are young versions of globular clusters.
Examination of the clusters from Cycle 1 {\it HST} images and early
post-refurbishment data
have already revealed that cluster A has two brightness peaks
and so is not as symmetrical as globular clusters are today 
(O'Connell \et\ 1994, De Marchi \et\ 1997).
What our data contribute to this continuing discussion, besides
the higher resolution compared to the earliest {\it HST} images, is
a deeper, and hence higher signal-to-noise, look at these clusters
and I-band images that are sensitive to the red supergiant
star population of the clusters.
In addition we examine the other compact clusters in the galaxy
to learn about the ensemble of star clusters formed in this 
unusual star formation episode in NGC 1569.

\section{The Observations and Data Reduction}

\subsection{The {\it HST} Data}

The center of NGC 1569 was imaged with the Wide Field and
Planetary Camera 2 (WFPC2) on {\it HST} on 1998 October 21.
The galaxy was centered on the PC for maximum resolution
of star clusters A and B: 0.04555\arcsec\ per pixel which is
0.55 pc at the galaxy. The galaxy was imaged through filters
F336W, F555W, F814W, and F656N. Exposures through F555W
and F814W consisted of a series of three integration times---short, medium,
and long. In the longer exposures the centers of clusters A and B
are saturated. For the longer exposures 
there are multiple exposures in order to improve signal-to-noise and
remove cosmic rays.
The list of observations are given in Table
\ref{tabobs}.

Basic data reduction steps were done by the Space Telescope Science
Institute ``pipeline'' processing system.
We produced a nebular emission image by combining the
medium-exposure F555W and F814W images, shifting, scaling,
and subtracting from the F656N image to remove the stellar
continuum.
We then subtracted nebular emission from the F555W and F814W
images, as needed, using a scaled and shifted \ha\ emission image. 
A mosaic of the CCD frames is shown in Figure \ref{figmos}.
A false-color combination of
F555W, F814W, and F656N is shown in Figure \ref{figcolor}.

We measured the brightnesses of field stars in the galaxy
using the crowded star photometry package DAOPHOT (Stetson 1987) as
implemented in the Image Reduction and Analysis Facility (IRAF). 
Here we will discuss the resolved stars in and near
the star clusters A and B; we will discuss analysis of the field star 
population itself in another paper. 

We
measured integrated photometry of star clusters using simulated
aperture photometry.
We calibrated the photometry using
the zero point constants given by Holtzman \et\ (1995b) in
their Table 9. We also converted the F336W, F555W, and F814W
photometry to the Johnson/Cousins UVI system using the 
conversions of Holtzman \et\ The instrumental magnitudes
were corrected for reddening and the red leak in F336W, as discussed
below, before
converting to UVI, as discussed by Holtzman \et

\subsection{Near-IR Spectroscopy}

We obtained longslit near-infrared spectra of clusters A and B
in order to examine the stellar CO features, which are an age
diagnostic through sensitivity to the presence of red supergiants. The spectra
were obtained over 3 nights in 1995 December 
with the Ohio State
Infrared Imager/Spectrometer (OSIRIS) on the 1.8 m Perkins telescope.
The spectra cover 21900 \AA\ to 23900 \AA\ at 8 \AA\ per pixel.
The clusters were observed in alternating positions on the slit,
10\arcsec\ apart. One position was used as a sky observation for the other.
We also alternated a 25--70 minute series of observations of a cluster 
with a pair
of observations of HR 1440, an A1V star without the CO spectral features,
chosen to be close to NGC 1569
in airmass. We also observed a series of
bright stars of various spectral types for comparison to the clusters.

We corrected the data for non-linearities using dark and flat field observations.
The flat sequence consisted of a series of increasing exposure times
intertwined with 1 second flat exposures to remove drifts in the lamps.
Dark observations preceeded and followed the flat field observations.
Regular dome flats were also taken and the data were flat fielded.
Sky frames were subtracted from object frames, 
one-dimensional spectra were extracted, and subgroups of spectra were combined.
Night sky lines in spectra with the sky not subtracted
were used to determine the wavelength scale and linearize
along the dispersion axis. The night sky lines were identified using a line list
provided by M.\ Hanson (private communication) which came originally
from C.\ Kulesa (private communication to M.\ Hanson).
The cluster spectra were divided by the nearest spectrum of HR 1440.
in order to remove the telluric absorption. For the cluster spectra
the continuum was also fit and the spectra divided by the fit.

\section{Data Analysis Issues}
\subsection{Reddening}

The reddening to and within NGC 1569 has been estimated by various
methods. 
The reddening in the Milky Way in the direction of NGC 1569 
E(B$-$V)$_f$ is estimated to be 0.51 by Burstein \& Heiles (1984)
from the column density of HI.
Israel (1988) measured a total reddening E(B$-$V)$_t$ of 0.56$\pm$0.10 
from ANS ultraviolet data. Devost, Roy, \& Drissen (1997) obtained
optical emission spectra of the ionized gas in NGC 1569 and concluded that
foreground E(B$-$V)$_f$ is 0.52 and internal E(B$-$V)$_i$ is 0.21 for R$=$3.1.
Using optical spectra Kobulnicky \& Skillman (1997) examined
the variation of reddening within the galaxy and found that
c(H$\beta$) varies from 0.8 to 1.2 with an average of 0.97$\pm$0.07.
The average translates into an E(B$-$V)$_t$ of 0.63 for R$=$3.1.

Figure \ref{figcolor} clearly shows that there is ionized,
and hence probably neutral gas and dust,
distributed non-uniformly around the galaxy.
From this figure one can see that 
clusters A and B themselves are sitting in a large hole
in the ionized gas and Israel \& van Driel (1990)
state that they are sitting in holes in the HI gas
as well (see also Swaters 1999). Thus, the extinctions to these clusters
are likely to be at the low end of the range in NGC 1569.
However, other clusters are embedded in \HII\ regions
and are likely to have extinctions at the high end
of the range. Nevertheless, according to Kobulnicky \& Skillman
(1997), the total range in E(B$-$V)$_t$ is only 0.56 to 0.71,
where 0.51 of that is foreground Milky Way reddening.

We have used Figure \ref{figcolor} to classify the clusters
in 3 extinction bins---heavy, medium, and light, according
to how much ionized gas is present in the immediate vicinity
of the cluster. Clusters 6, 41, and 44 are considered to be
heavily internally extincted and E(B$-$V)$_t$ is taken to be 0.71 for these.
Clusters 4, 7, 8, 9, 10, 39, 40, and 42 are considered
to have medium extinction and E(B$-$V)$_t$ is taken to be
0.63. All other clusters are taken to be lightly internally extincted
and an E(B$-$V)$_t$ of 0.56 is adopted. 
Although this manner of determining the reddening is rough,
the uncertainty in any adopted E(B$-$V) can be no more than
$\pm$0.15 magnitude, the total range in E(B$-$V) measured
from spectroscopy.
We use the Cardelli, Clayton, \& Mathis (1989) extinction curve and an 
A$_V$/E(B$-$V) of 3.1. (Although NGC 1569 is of low metallicity
compared to the Milky Way,
this has small effect on the optical extinction curve [Bouchet
\et\ 1985]).

Holtzman \et\  (1995b) showed that the reddening correction is
a function of the spectrum of the object, and at F336W the difference
between the extinction of an O6 star and a K5 star can be large.
The colors of the clusters cover a large range, comparable to that
of the stellar main sequence. Therefore, we have determined a reddening
correction that depends on the integrated colors of the clusters.
To determine the extinction as a function of the observed colors
of objects, we used the tools in the Space Telescope Science Data Analysis
System (STSDAS) to simulate the
throughput of the telescope plus filters for various blackbodies
reddened by E(B$-$V)$_t$'s of 0.56, 0.63, and 0.71. 
The extinction corrections for O6 and K5 type spectra are those
given by Holtzman \et\ in their Table 12; we have simply used 
STSDAS simulations to determine extinctions for effective temperatures
in between these.
The cluster photometry
was then corrected for extinction according to its observed
F555W$-$F814W color and its E(B$-$V)$_t$ category. 

\subsection{Red Leak in F336W}

Since we wish to examine the F336W$-$F555W color of clusters,
we need to consider the red leak in the F336W filter. Not only
are many of these clusters intrinsically red, but they are also
highly reddened from foreground extinction. This means that
red photons could contribute significantly to the counts in
the F336W filter.
To determine the red leak as a function of the
observed F555W$-$F814W color, we used the simulations in STSDAS
and blackbody curves reddened by E(B$-$V)$_t$'s of 0.56, 0.63, and 0.71.
The red leak was taken to be any flux contribution from $\geq$4000 \AA,
after the definition of Holtzman \et\ (1995b).
The F336W cluster photometry was corrected for the red leak
based on its observed F555W$-$F814W color and its reddening
category. For an especially red cluster with an observed
F555W$-$F814W of 1.5
and an E(B$-$V)$_t$ of 0.71,
the contribution of the red leak to the F336W magnitude is 0.06 
magnitudes, but the correction increases rapidly
for redder observed colors.

\subsection{Distance to the Galaxy}

Israel (1988) used a distance to NGC 1569 of 2.2$\pm$0.6 Mpc.
From WFPC1 observations O'Connell \et\ (1994) measured
the brightnesses of the brightest stars in NGC 1569,
and, assuming an E(B$-$V)$_t$ of 0.56 from Israel (1988),
determined a distance of 2.5$\pm$0.5 Mpc. Greggio \et\
(1998) examined the color magnitude diagram of the galaxy
from WFPC2 images and adopted an E(B$-$V)$_t$ of 0.56
and distance of 2.2 Mpc without apparent distress
to the analysis of the stellar population.
In this paper we will adopt 
a distance of 2.5 Mpc.

\section{Identification of Clusters}

Clusters A and B are the most spectacular star clusters in NGC 1569,
but they are not the only ones. We have examined the WFPC2 images
for other compact, but less luminous, clusters. We have included
as a cluster any compact object that was resolved compared to 
an isolated star profile. Contamination of the sample by background galaxies is
possible although WF2 and WF3, which include little of NGC 1569,
contain few background galaxies and no objects outside of NGC 1569
that fit our criteria.

Generally, we have not included
looser, and spatially bigger, OB associations in our list of clusters. 
The reason for this is
that the galaxy is really one big OB association and separating one
from another is not feasible. Also, we are interested in the less common
compact clusters for comparison with clusters A and B. However, we did
include one OB association, cluster number 48 on WF4. This OB association
is discussed by Drissen \& Roy (1994) who discovered a ring nebula
surrounding the cluster and broad stellar emission lines that they
attribute to a WN star. It is located in the outer part of NGC 1569
and was easy to isolate. It serves as an example of a certified
OB association for comparison to the other clusters.

Because the clusters are so numerous, we could not reasonably continue
the identification scheme begun by others with clusters A and B. Therefore, 
we switched to using numbers as identifiers, but began with number 3,
so clusters A and B are also numbers 1 and 2.
The clusters are identified in Figure \ref{figclus},
and are listed in Table \ref{tabinteg}.

Ten of the newly identified clusters are located in the vicinity of 
super star cluster A, over a region of about 50 pc projected on the sky.
Together they may form something akin to a small version of the 
greater 30 Doradus
region in the LMC. 30 Doradus is a kpc-sized area in which star formation has
taken place in small units here and there over a timescale of order 10 Myrs
(Walborn \& Blades 1997; see also Constellation III, Dolphin \& Hunter
1998).
Otherwise the clusters are distributed throughout the galaxy
with no apparent pattern.

\section{Integrated Cluster Photometry}

We give integrated photometry of the star clusters in
Table \ref{tabinteg}.
The contributions from the sky and background galaxy were measured in an annulus
just beyond the integration aperture for each object. The aperture radius
is given in Table \ref{tabinteg}.

We show the integrated colors and magnitudes in color-color and
color-magnitude diagrams in Figures \ref{figclccd} and \ref{figclcmd}.
We also include cluster evolutionary tracks from Leitherer \et\ (1999).
We use their tables for an instantaneous burst of star formation
and a Salpeter (1955) stellar initial mass function
from 1 to 100 M\solar. Time steps of 1 to 9 Myrs in steps of 1 Myrs
are marked with an ``x'' along this evolutionary
track; time steps of 10, 20, and 30 Myrs are marked with open circles.
The M$_V$ have been adjusted from the models of 10$^6$ M\solar\
to the mass of 3.3$\times10^5$ M\solar\ determined for cluster A
by Ho \& Filippenko (1996). For less massive clusters the tracks
in Figure \ref{figclcmd} would slide vertically to fainter M$_V$.

Kobulnicky \& Skillman (1997) give the oxygen abundance of emission
nebulae in NGC 1569 as
8.19$\pm$0.02 with no evidence for chemical inhomogeneities.
This oxygen abundance implies a metallicity $Z$ of approximately
0.004. One would expect that the metallicity of current \HII\ regions
in NGC 1569 should be a good estimate of the metallicities of the recently
formed stars. 
Therefore, we began with the $Z = 0.004$ models of
Leitherer \et\ (1999). However, we found that the $Z = 0.004$
cluster evolutionary tracks did not account very well for the colors
of some of the clusters, but that $Z = 0.008$ models did. 
Therefore, in Figures \ref{figclccd} and \ref{figclcmd} we
include cluster models for both $Z = 0.004$ and $Z = 0.008$.

In Figures \ref{figclccd} and \ref{figclcmd} we see that cluster
A has UVI colors that are consistent with an age of order
4--5 Myrs, according to cluster evolutionary models for both metallicities.
This is consistent with the detection of the signature of
Wolf-Rayet stars in the cluster (Delgado \et\ 1997).
However, O'Connell \et\ (1994) and De Marchi \et\ (1997) have shown 
that cluster A has two peaks
in the light distribution, suggesting two sub-clusters. De Marchi \et\
also suggest that these two sub-clusters may have different ages,
with one being dominated by Wolf-Rayet stars and the other dominated
by red supergiants.
However, the integrated UVI colors of this cluster do not reflect that
and appear to be dominated solely by the younger component.

Many of the other clusters in Figure \ref{figclccd} likewise fall near the
locus of the $Z = 0.004$ models.  The group lying above cluster A
in the diagram and slightly blueward of the locus can be explained
if we have made a small overcorrection to their extinction.

However, cluster B and 8 fainter objects are too red in (V$-$I) to
be consistent with the $Z = 0.004$ locus.  We believe this is a
symptom of an important contribution from red supergiants to the
U,V,I light.  The colors of the Leitherer \et\ (1999) models
exhibit strong sensitivity to abundance in the range $Z = 0.004$
to $Z = 0.008$ for ages 8--16 Myr, as shown in Figure \ref{figclccd}, 
because of the
influence of red supergiants at these ages.  Unfortunately, as
emphasized by Origlia \et\ (1999), there is
considerable uncertainty about the evolutionary tracks for such
phases at subsolar metallicities.  What we can say is that
cluster B and the other objects in its vicinity in Figure \ref{figclccd} are
roughly consistent with the $Z = 0.008$ models of Leitherer et al. (1999)
at an age betweeen 8 and 15 Myr. The agreement for cluster B 
would be particularly 
good if we have overcorrected 
for extinction by about 0.2 mags in E(B$-$V). However, the E(B$-$V) that
we assumed for cluster B is only 0.05 magnitude above the estimated foreground
extinction, so it is unlikely that we have overestimated the extinction
by very much.
An age of 8--15 Myrs, however, is
consistent with the CO-band evidence for red supergiants
discussed in Section 6.3. 

The youngest cluster, occupying the upper left in Figure \ref{figclccd},
is number 48, the OB association in the southeast part of the galaxy.
Its colors indicate an age of just 2--3 Myrs. This age is consistent with
the upper limit of 5 Myrs placed by Drissen \& Roy (1994) based on
their observation of an expanding bubble due to a WN-type evolved
massive star.
The rest of the clusters, with the exception of the group around
cluster B, span the range of colors along the cluster evolutionary
sequence.

Part of the scatter seen in Figure \ref{figclccd} is caused by
stochastic effects. The latter are discussed and simulated by Girardi
\& Bica (1993; see also, Santos \& Frogel 1997; Brocato \et\ 1999). 
Stochastic effects are due to small number
statistics in populating the upper masses of the stellar
initial mass function; as the few massive stars in small clusters
evolve, the status of a single star can have a profound affect
on the integrated colors of a small cluster. Girardi and Bica's
simulated color--color diagram shows a scatter in UBV colors
of several tenths magnitude.
This problem affects clusters as old as 1 Gyr as well as young ones
(see, for example, Gallagher \& Smith 1999).
Brocato \et\ further point out the difficulty
in correcting for the red leak in {\it HST} filters under
these circumstances. In addition, differences of
several tenths magnitude in colors can be seen between different
models depending on the treatment of various evolutionary parameters
(see, for example, Brocato \et\ 1999).
Thus, there is considerable uncertainty in assigning an age
to a specific cluster from global colors.

Keeping the observational, stochastic, and modeling
uncertainties in mind, we conclude from Figure \ref{figclccd}
that the clusters
roughly span the full range in ages covered by the evolutionary
models from the OB association, which is the youngest at a few
Myrs, to nearly 1 Gyr.
However, the distribution is very unlike
what would be expected for a uniform distribution of ages.
Instead, there is a very strong concentration to ages less than
30 Myr, with a subconcentration at 4--6 Myr. 
Therefore, most of the clusters have been formed in the recent
starburst activity of the galaxy. The presence of a few potentially older
clusters suggests that the galaxy also formed stars in compact
clusters even before the
advent of the present starburst.

Clusters A and B are by far the most luminous of the clusters.
The rest of the clusters in our sample
have M$_V$ between $-$7 and $-$12.
As shown in Figure \ref{figclcmd}, the model
clusters evolve at about constant M$_V$
for the first 7 Myrs, and M$_V$ steadily
declines after that, losing several magnitudes by the time the
cluster reaches the colors of the reddest clusters in our sample.
Therefore, the bluer clusters with M$_V$ near $-$11 and the redder clusters
with M$_V$ near $-$8 are likely to have had absolute magnitudes
of order $-$11 when they were a few Myrs old. 
This implies that these clusters are comparable in stellar mass
to the compact cluster R136 in the LMC. R136 has an M$_V$ of
$-$11 at an age of 2 Myrs (Hunter \et\ 1995, Massey \& Hunter 1998).
Since R136 is comparable to a small globular
cluster in mass and compactness, it appears that NGC 1569, besides
making two very luminous super star clusters, has also made another half
dozen or so clusters comparable to small globulars as well. 

In addition to the more luminous star clusters, however, there are 
many less luminous clusters.
Those with M$_V$ near $-$7 likely contain $\leq$15 massive
stars, and would be considered more like OB associations in
terms of stellar content if they were not so compact in size.
The OB association, number 48, included in the sample of clusters
has an M$_V$ of $-$8.8 and was encircled by an aperture with
radius 14 pc. Except for clusters A and B, the other clusters
were observed with apertures smaller than this: 0.2--11 pc,
with an average of 3 pc, in radius.
Some of these
may be comparable to the LMC's populous clusters.
For example, the populous cluster NGC 1818 in the LMC
has an M$_{F555W,0}$ of $-$9.3 now and probably $-$10.5 at an age
of 4 Myrs with a half-light radius of 3.2 pc (Hunter \et\ 1997).

Thus, it appears that NGC 1569 in its recent burst of star formation
has formed many compact and relatively massive star clusters.
These clusters include the extreme super star clusters A and B,
as well several clusters comparable to small globular clusters
in luminosity and many other comparable to normal OB associations
but more compact than is typical.
We do not understand what conditions are necessary to produce
compact clusters or super star clusters (see, for example,
discussion by Dolphin \& Hunter 1998). However, those conditions
appear to have been met recently in NGC 1569.

\section{The Super Star Clusters}

From {\it HST} Cycle 1 data, the super star clusters A and B were found to have 
half-light radii \rhalf\
of 2.2 and 3.0 pc, compared to \rhalf\ of 1--8 pc for today's Milky Way
globular
clusters (van den Bergh \et\ 1991). The integrated V-band magnitudes
of the clusters in NGC 1569 are
$-$14 and $-$13, corrected for extinction.
As shown in Figure \ref{figclcmd}, both would have had M$_V\sim-14$
at an age of 4 Myrs. 
By contrast, a typical globular
cluster, if it had a Salpeter (1955) stellar initial mass function, would
be expected to have had a magnitude of $-$13.7$\pm$1.3 (Harris 1991).
Thus, the clusters in NGC 1569 have luminosities and sizes
that are comparable to globular clusters
that are seen in our Galaxy today, and they are excellent candidates
for being true young globulars.

The extreme nature of
NGC 1569's super star clusters have long intrigued astronomers,
and some properties, particularly their ages, have remained controversial.
Ho \& Filippenko (1996) used velocity dispersion data to estimate
a mass of 3.3$\pm$0.5 $\times10^5$ M\solar\ for cluster A.
Sternberg (1998) combined photometric data with a velocity dispersion
and concluded that in cluster A there are stars present down to the
hydrogen burning limit of 0.1 M\solar.
Arp \& Sandage (1985) obtained integrated spectra of the clusters
and found that they have spectra like those of A supergiants.
Prada, Greve, \& McKeith (1994) obtained spectra of the Ca II infrared
lines of the clusters and concluded, as had Arp and Sandage, 
that cluster A is dominated by A and B stars and is 13--20 Myrs old.
Cluster B, they concluded, is in a red supergiant phase and is 12 Myrs
old. 
O'Connell \et\ (1994) estimated the ages of both clusters 
to be roughly 15 Myrs from
integrated V$-$I colors obtained from {\it HST} images.
In optical ground-based 
spectra Delgado \et\ (1997) detected the signature of Wolf-Rayet
stars in cluster A as well as the signature of young, massive stars
in both clusters. They concluded that each cluster had undergone
two episodes of star formation separated by about 6 Myrs: 3 and 9 Myrs
in cluster A and 2 and 8 Myrs in cluster B.
However, De Marchi \et\ (1997) from {\it HST} images concluded that
cluster A is actually two clusters, one that is dominated by red supergiants
and one that is younger and contains the Wolf-Rayet stars.

Here we examine some additional data on the structures and ages of
these super star clusters.
In addition we will include in some of the analysis below two
other clusters from Table \ref{tabinteg} for comparison.
The comparison clusters
were chosen to be well-resolved and among the brightest few clusters
after clusters
A and B: Cluster 30 has an M$_V$ of $-$11 and cluster
35 has an M$_V$ of $-$10.

\subsection{Structure}

In Figure \ref{figcontour} we show contour plots of clusters
A, B, 30, and 35.
We can see from Figure \ref{figcontour} that cluster A
has two peaks in its two-dimensional profile, as first
pointed out by O'Connell \et\ (1994) and later analyzed
De Marchi \et\ (1997). The peaks, designated A1 and A2 by
De Marchi \et,
are separated by 0.18\arcsec, which is 2.2 pc at the galaxy,
and the fainter peak is southeast of the brighter peak.
Clusters B, 30, and 35, on the other hand, are relatively
round. In the outer parts, cluster B is not
entirely symmetrical, being elongated in
one dimension, but the deviation is small.

Because of interest in structure within clusters A and B, we have 
produced two-dimensional ratio maps of these clusters using
the F555W and F814W images. We subtracted local background galaxy
from each cluster in each image, and aligned all of the images
to the medium F555W exposure.
We combined the short, medium, and long exposures, replacing the saturated
pixels with values from the unsaturated exposures.
The resulting ratio images are shown in Figure \ref{figratio}.
One can see that both
clusters are relatively blue in the centers. 
In cluster A the two luminosity peaks
A1 and A2
are very similar in color and both are blue. The reddest regions
are located in the outer parts of cluster A and are not obviously
associated with any one peak. 
This is not what one would expect if one component is dominated by
red supergiants and the other is not as suggested by De Marchi \et\ (1997).
However, the large overlap of the
two components makes it hard to disentangle the outer parts of the two
sub-clusters.
Cluster B, on the other hand, is more uniform in appearance,
and the center is not as blue although there is a bluer clump of stars
to the southeast of the center.

We have also experimented with deconvolving components A1 and A2 in
cluster A, as had De Marchi \et\ (1997). 
We used point-spread-function deconvolution (DAOPHOT, Stetson 1987)
with various combinations of clusters B, 30, and 35 
with Gauss, Lorentz, Penny, and Moffat fitting
functions to determine the point-spread-function. 
We also fit two two-dimensional Gaussians to the peaks.
Residuals were fairly high in all cases arguing that clusters B, 30,
35, and a two-dimensional Gaussian do not well represent the structures
of A1 and A2. Nevertheless, the derived parameters are a rough indication
of the characteristics of the two components.
We find that component A1 is brighter than A2 by about 1.7$\pm$0.2 magnitudes
in F555W. De Marchi \et\ (1997) found a difference of 1.3 magnitudes.
Component A1 has an (F555W$-$F814W)$_0$ color of 0.13. A2 is 0.06
magnitude redder than A1 in these filters, but within the uncertainties
one could also say that A1 and A2 have the same colors.
De Marchi \et\ found A2 to be 0.1 magnitude bluer in (F439W$-$F555W)$_0$.
From the two-dimensional Gaussian fits to the F555W image we find that A1 is
2.5$\times$1.9 pc 
(0.21$\pm$0.01\arcsec$\times$0.155$\pm$0.007\arcsec)
FWHM at a position angle of 35\arcdeg, and
A2 is 1.9$\times1.4$ pc 
(0.15\arcsec$\pm$0.01$\times$0.12$\pm$0.02\arcsec)
FWHM at a position angle of 105\arcdeg.
De Marchi \et\ measure a half light radius of 0.15\arcsec\ for A1
and of 0.17\arcsec for A2.

\subsection{Integrated and Surface Brightness Profiles}

The surface brightness and integrated radial profiles
of clusters A, B, 30, and 35 are shown in 
Figure \ref{figprofile}. 
The annuli were measured in steps of 2 pixels which is 0.09\arcsec.
This choice of step size is small enough with respect to
the size of the cluster to give interesting radial information.
However, it is still much larger than uncertainties in 
aligning the images. Uncertainties due to variations in
the point-spread-function between filters
are expected to be less than 1\% for radii $\geq$4 pixels,
but the inner most point in radial profile plots will be
affected especially F555W$-$F814W (Holtzman \et\ 1995a).
To increase the signal-to-noise,
we found it necessary to integrate
over a 4-pixel wide annuli for the annular color profiles.

The fluxes lost due to the saturated pixels 
in the images of clusters
A and B in the medium and long F555W and F814W exposures have been
corrected using the short exposures. The final aperture photometry 
is the
average of the photometry in the three exposures, and the uncertainties
are the dispersion around the mean. 
These are observed magnitudes, and F336W has not been corrected for red leak here.

In terms of colors most of the clusters are fairly uniform with radius.
All of the clusters are bluer at the center in F555W$-$F814W by 0.05--0.1
magnitude, but this is probably due to the difference in point-spread-function
between these two filters (Matthews \et\ 1999).
The F555W$-$F814W color then
quickly becomes redder, reaching a level that does not change significantly
with radius.
In F336W$-$F555W, the behaviour is somewhat different. F336W$-$F555W remains
fairly constant with radius in clusters A and 30. On the other hand,
F336W$-$F555W in clusters B and 35 becomes steadily bluer with radius,
by of order 0.1 magnitude. 

The half-light radii \rhalf\ in F555W are given in Table \ref{tabking}.
They are 2.2--3.1 pc for clusters A, B, 30, and 35. 
Our measured values for clusters A and B
are very close to those measured by O'Connell \et\ (1994) from
Cycle 1 {\it HST} data. 
A typical value of \rhalf\ for globular clusters is 3--4 pc
with a range from 1--8 pc (van den Bergh, Morbey, \& Pazder
1991), assuming no dynamical evolution of the clusters.
Thus, clusters A, B, 30, and 35 have half-light radii that
are nearly typical of globular clusters. 
Thus, clusters A and B have the luminosities (discussed 
in Section 5)
and \rhalf\ of a typical Milky Way globular cluster. 
In addition,
Ho \& Filippenko (1996) have determined the dynamical mass of cluster A
to be 3.3$\pm$0.5$\times10^5$ M\solar. An average mass of a globular
cluster is 6$\times10^5$ M\solar with a range in masses of
10$^4$ M\solar\ to 4$\times10^6$ M\solar (Pryor \& Meylan 1993).
Thus, cluster A clearly has many attributes of
a young version of a globular cluster.

Clusters 30 and 35 are the reddest clusters in our sample and, with
M$_V$ of $-$11 and $-$10, are several magnitudes fainter than A and B. 
In Figure \ref{figclccd} they lie near the 1 Gyr mark
of the $Z = 0.004$ cluster evolutionary track. If they are truely that old,
then at 10 Myrs they would have been comparable to cluster A in luminosity,
implying that super star cluster formation has occurred at various
epochs in the past. Because we have assumed a light reddening for these clusters,
it is unlikely that we have overcorrected for reddening and that they are 
consequently really bluer.
However, it is possible that the colors of clusters 30 and 35 are dominated
by a few red supergiant stars.
If clusters 30 and 35 are only 30 Myrs
old, as stochastic uncertainties might allow, 
their M$_V$ at 10 Myrs
would have been $-$12 and $-$11, still placing them in the regime
of smaller globular clusters and of R136 in the LMC.
Spectra would be necessary to distinguish between these possibilities.

We have fit the surface brightness profiles of clusters B, 30, and 35
with a King model (King 1962). Cluster A is excluded from this analysis
because it is clearly not spherically symmetric. Following the proceedure
outlined by King, we used the inner and outer parts of the profile to
determine first guesses of key parameters. 
We plotted R-squared against the inverse of the
flux and fit the inner part of the profile to determine the core
radius r$_c$ and the central flux f$_0$. We plot 1/r against the square-root
of the flux and fit the outer part of the profile to get an estimate
for the terminal radius r$_t$. With these first guesses, we then
explored the parameter space of reasonable solutions, determining
the best fit and the range of values that are reasonable for these three
variables. The uncertainties
in the former are determined from the latter. The best fit for each
cluster is shown as the solid line in Figure \ref{figprofile}.
Fit parameters are listed in Table \ref{tabking}.

We have used the list of structural parameters for present-day Milky Way globular
clusters given by Djorgovski (1993) and Trager, Djorgovski, \& King (1993)
to compile a range and average of the core radius r$_c$ and the
concentration index c. We included globular clusters with data quality
indices given by the authors of 1 or 2 and excluded clusters identified
as having undergone core collapse. This yields an average core radius
for 89 globular clusters of 2.6$\pm$3.9 pc with a total range of 0.2--23 pc.
NGC 1569's clusters B, 30, and 35, by comparison, have core radii
of 0.3--0.8 pc. The core radii of NGC 1569's clusters
are smaller than that of the average Milky Way globular cluster,
but within the range that is observed.
For 95 Milky Way globular clusters the average concentration index
is 1.4$\pm$0.4, with a total range of 0.6--2.4. 
The concentration index of the clusters in NGC 1569 are
1.7--2.0. The clusters in NGC 1569 have concentration indices that
are within range of
what is observed in Milky Way globulars today but larger, that is
less concentrated, than the average.

Thus, cluster A in 
NGC 1569 has the half-light radius, luminosity, and mass of a
fairly normal globular cluster, as seen in the Milky Way today. However,
unlike globular clusters seen today, it is not spherically 
symmetric like the Milky Way globulars or elliptically symmetric like
the LMC globulars. However, the globular clusters may have
been rounded-out over time by the tidal forces in the Milky Way 
(Goodwin 1997). Cluster A has two luminosity peaks that could
be sub-clusters.
Perhaps the two components will merge in the future.
Cluster B also has the half-light radius and luminosity of a
typical globular cluster. In addition its core radius is at the
small end of the range of globulars and it is not as concentrated.
Clusters 30 and 35 have half-light radii, core radii, and luminosities that
are a bit smaller than the average of the globulars but within
the range observed. However, they are not as concentrated overall as the
average globular.

\subsection{The Stellar CO Feature}

The near-infrared stellar absorption features of CO have been found to
be useful for placing limits on the age of a coeval stellar population.
The features around 2.3 microns are found in stellar
spectra beginning with late G stars and becoming stronger with later
spectral type of the star (Baldwin \et\ 1973). Thus, in a young cluster
the CO features are sensitive to the presence of massive stars
entering the red supergiant evolutionary phase. Prior to the onset
of the red supergiant phase, a young coeval cluster will exhibit
no CO features, and then as the red supergiants appear the CO features
appear in the integrated cluster spectrum.
For the metallicity of NGC 1569, this occurs
at $\sim$7 Myrs (Mayya 1997, Origlia \et\ 1999). 
While the red supergiant peak only lasts $\sim$5--7 Myrs 
(Bica \et\ 1990, Origlia \et\ 1999), after that
other red stars continue to contribute to the CO feature, including
an AGB phase at $\sim$100 Myrs (Olofsson 1989, Bica \et\ 1990). 
Thus, the CO features can be used to set an upper or lower limit 
on the age of the clusters depending on whether the features are
present or not.

Therefore, we undertook to measure the near-infrared stellar CO absorption
features in the super star clusters in NGC 1569.
The near-infrared spectra of clusters A and B are shown in Figure \ref{figco}.
The $^{12}$CO(2,0) and $^{12}$CO(3,1) 
stellar absorption features around 2.3 microns
are prominent. The equivalent widths of these features are given in
Table \ref{tabco} for the clusters and other stars that were observed
for comparison.
For cluster A this measurement integrates over
both sub-clusters (O'Connell \et\ 1994, De Marchi \et\ 1997).
One can see that the equivalent widths of the stellar CO spectral
features in clusters A and B are very strong. They are, in fact,
comparable to the equivalent widths of those features in
the spectra of K and M
giants and supergiants.
Uncertainties in the equivalent widths are not given because the
uncertainties are dominated by systematics. From the observation
of the F7V star, that should have an equivalent width of zero,
we estimate that the uncertainty in the
equivalent widths is of order 2 \AA.

The large CO equivalent widths in the super star clusters indicate
the presence of a significant population of cool, luminous
stars, though without further information we cannot determine the
exact types. The dilution of the CO absorption features by the light
of warmer stars means that the dominant cool types have equivalent
widths $>$30 \AA\ and are probably red supergiants.
Because both clusters show the strong signatures of red supergiants, 
then, their ages
must be $\geq$7 Myrs. The integrated colors, as shown in Figure \ref{figclccd}
suggest that cluster A is at the low end of this limit, while cluster B
is perhaps somewhat older. 

Prada \et\ (1994) observed the Ca~II near-infrared
spectral features in clusters A and B. 
They concluded that cluster B is in the red supergiant
phase but that cluster A, with weaker Ca~II, is older.
However, the Ca II calibrations of Bica \et\ (1990) do not
preclude a younger age for cluster A. As discussed above, 
the integrated colors of cluster A point to a younger age.

\subsection{Resolved Stars}

Photometry of resolved stars in the outer parts of clusters A and B
are shown superposed on color--magnitude diagrams of the entire
measured field star population in Figures \ref{figcmdcl}. 
The stars are shown as a function of distance from the center of the
clusters in Figure \ref{figclradius}.

As anticipated by O'Connell \et\ (1994), a large component of the
resolved stars in and near clusters A and B are red supergiants.
These are stars with M$_V\leq-4$ and V$-$I$\geq1.5$.
In cluster A there are approximately 8 stars with the colors and absolute
magnitudes of red supergiants.
These are all located along the edge of the included region around the cluster.
The rest, and the majority, of the resolved stars are 
very bright stars in the blue plume
extending from about 23.5 to 19 in F555W. This corresponds to
$-$5.2 to $-$9.7 for an E(B$-$V)$_t$
of 0.56. 
Early O-type supergiants have M$_V$ of order $-$7, and we do see a large clump
of blue stars between $-$5.5 and $-$7 in cluster A. 
These are likely massive O stars.
The color-magnitude diagram of cluster A, therefore, 
is consistent with the interpretation
of other data that there are two stellar populations present
or a spread in ages (Delgado \et\ 1997).

Of the resolved stars in cluster A, 16 have M$_V$$\leq$$-$7,
even brighter than an O supergiant (Conti \& Underhill 1988).
What are these brightest stars? 
A possibility is that they are clumps of stars although there would have to 
be as many as 10 stars in some cases,
but a more intriguing,
and equally likely,
possiblity is that they are evolved massive stars. For example,
Luminous Blue Variables (LBVs), which come from stars $\geq$85 M\solar\
(Massey, Waterhouse, \& DeGioia-Eastwood 2000), 
can be very bright in V: S Doradus in the LMC
is currently $-$9.7 (Massey 2000). Yet, because they are cooler,
LBVs can be brighter than O supergiants or Wolf-Rayet stars.
Furthermore, LBVs
have roughly the same ages as Wolf-Rayet stars, which come
from stars more massive than about 30 M\solar\ (Massey, Waterhouse,
\& DeGioia-Eastwood 2000), and so their 
presence is consistent
with the presence of 20--40 WN-type stars in the cluster, estimated by 
Delgado \et\ (1997).
Cluster A is comparable to a globular cluster. Therefore,
it also likely contains millions of stars and, thus, as in R136, we expect
there to be a considerable massive star population in the cluster, including
numerous of the most massive stars known
(Massey \& Hunter 1998). For ages of order 4--5 Myrs,
these very massive stars will not yet have exploded as supernovae. 
The presence of so many Wolf-Rayet stars confirms
this picture, and since a large number of Wolf-Rayet stars are present, 
it would not
be a surprise to find LBVs too.

The resolved stellar population in and near cluster B, on the other hand, 
is markedly 
different in make-up compared to that in cluster A.
There is a clump of bright, blue stars. These are comparable in 
magnitude to O supergiants 
and main sequence stars. The very brightest blue stars seen in cluster A are not
present in cluster B.
However, the majority of the stars that we have resolved 
are red supergiants. The large number of red supergiants are consistent with
the strong stellar CO feature and the conclusion that cluster B is in its 
red supergiant phase.

\subsection{Ages}

So, where does this leave us concerning the ages of clusters A and B?
For cluster A the integrated colors and the presence of Wolf-Rayet stars argue
for ages of order 4--5 Myrs, and the presence of red supergiants
and the stellar CO equivalent width argues for
an age greater than 7 Myrs.
This apparent
contradiction between a young age and an older age is the reason
that De Marchi \et\ (1997) suggested that one of the sub-clusters
in A is the young object and the other is the older object. 
However, with this scenario one might expect some spatial separation in stellar
populations and hence colors.
While plausible, our two-dimensional color map does not offer support
of this. Alternatively, cluster A may have formed over an extended period
of time, as suggested by Delgado \et\ (1997), 
which would potentially allow the older and younger stars
to be more spatially mixed.
An age spread of only a few Myrs in the formation of the massive stars
is all that would be required.
Furthermore, Massey (1998) and Massey \& Johnson (1998) have shown
that RSG and Wolf-Rayet stars are often found together in OB associations.
The same mass range of massive stars go through the RSG and the
Wolf-Rayet phases, and this is especially true for a metal poor stellar population. 
Thus, RSGs and Wolf-Rayet stars can coexist
even in a coeval population.

The data for cluster B, on the other hand, are more consistent.
The colors suggest 10--30 Myrs, with uncertainty because the
cluster does not fall near the cluster evolutionary track, the large number
of red supergiants and the CO equivalent
width argues for an age greater than 7 Myrs, and the Ca II feature
also suggests that the cluster is in the red supergiant
phase (Prada \et\ 1994).
Thus, it appears that cluster B is of order 10--20 Myrs old and,
therefore, older than the young component of cluster A, but 
roughly comparable in age to the older component of A.

According to Vallenari \& Bomans (1996) and 
Greggio \et\ (1998), the bulk of the current starburst in NGC 1569 ended
5--10 Myrs ago and extended back about 100 Myrs from that.
It is expected that super star clusters might form during such an episode,
although it is surprising that cluster formation was so strongly
concentrated to the end of that period. 
Once clusters form, they are expected to disperse the gas relatively
quickly (as seen in Figure \ref{figcolor}).
According to Israel \& van Driel (1990), the size of the HI hole---diameter 
200 pc---around
the clusters is consistent with an age of 2--10 Myrs
for reasonable input parameters to stellar wind models.
The hole in the \ha, seen in Figure \ref{figcolor}, is 170 pc
in diameter and is also consistent with this picture.
In spite of the obvious disruption of the ISM in the center of the galaxy,
star formation is, nevertheless, continuing: 
Taylor \et\ (1999) have mapped 5 giant
molecular clouds just outside the HI hole 
to the northwest of the two super star clusters.

\section{Summary}

We have examined star clusters in {\it HST} images of the irregular,
starburst galaxy NGC 1569. In addition to the super star clusters A and B that
are well known, we identified 45 other clusters that are compact but resolved.
We also include one known and easily isolated OB association in our study.
The OB association is the youngest of the clusters.

Integrated colors of the clusters span a large range. Comparison with
coeval evolutionary models at $Z = 0.004$ 
suggest that the ages range from a few Myrs to
nearly 1 Gyr, but a better fit to some cluster colors may be obtained with
more metal-rich $Z = 0.008$ models.  
Uncertainties due to stochastic effects, reddening, background
subtraction, and modelling
make it difficult to pin ages of clusters down from integrated photometry.
However, a large number of the clusters have colors consistent with ages of
4--6 Myrs and most with ages $<$30 Myrs.
Thus, most of these clusters have been formed at the tail end
of the recent burst of star formation and were not formed continuously
during the starburst.

We examined the surface brightness profiles of the super star clusters A and B
and two other very bright clusters. We fit King model profiles to all 
but cluster A.
The clusters have half-light radii and core radii that are in the range observed
for present-day globular clusters in our Galaxy. 
However, the core radii of the NGC 1569 clusters are on the small
end of the range and they are somewhat less concentrated that the average
globular. Clusters A and B have luminosities, and cluster A has a mass,
that place them as young versions of typical globular clusters. The other
two clusters, and likely several others in the sample as well, are similar
to a small globular cluster and to R136 in the LMC.
Thus, whatever conditions facilitated the recent starburst were also adequate
for producing numerous compact clusters that are comparable to small
globular or populous clusters.

Cluster B and two other clusters chosen for comparison are nearly spherically
symmetric. However, cluster A, as was known from previous work,
has two peaks in its light distribution. This has been interpretted by
others as evidence for the presence of two sub-clusters.
The colors of the clusters are approximately constant with radius,
with F336W$-$F555W becoming bluer by 0.1 magnitude in two of the clusters.

We have examined resolved stars in the outer parts of clusters A and B.
We find that cluster A contains many
bright blue stars. Some of the blue stars are bright enough to be
evolved massive stars such as LBVs. The presence of such stars is consistent
with the presence of a large number of Wolf-Rayet stars and the expected
large number of very massive stars for the mass of the cluster.
The resolved star population around cluster A also contains a small
population of red supergiants. The presence of both 
blue and red massive star populations are
consistent with the view that the two sub-clusters have different
ages, but
our two-dimensional color map of cluster A suggests that
the two luminosity peaks are comparably blue and that the reddest stars,
located in the outer parts of the cluster, are
not obviously preferentially associated with one of the peaks.
Instead, the presence of both RSG and Wolf-Rayet stars in a metal poor coeval
population is not inconsistent with the evolution of massive stars, 
or alternatively there is a small age spread 
of several Myrs within the cluster.
The stars that we resolved around cluster B, on the other hand,
contain a small population of more normal blue massive stars and
a large population of red supergiants. The presence of the red
supergiants is consistent with the view that cluster B is in
its red supergiant phase.

We have measured the near-infrared stellar CO feature in clusters A and B
and find that strong absorption features are present in both. This indicates
that red supergiants are present, as is confirmed by photometry of the 
resolved stars in
the outer parts of the clusters. The implied ages are $\geq$7 Myrs.
Other age indicators are consistent with a picture in which cluster B
is of order 10--20 Myrs old.
It is likely
that the clusters have contributed to clearing a hole in the gas,
seen in HI and \ha. The \ha\ hole is 170 pc in diameter,
and the timescale to form it is comparable to the age of
cluster B. 

\acknowledgments

We wish to thank Allan Watson for help with the near-infrared observations.
Support for this work was provided by NASA through
grant number GO-06423.01-95A to D.A.H. and grant number
NAG5-6403 to R.W.O. 

\clearpage

\begin{deluxetable}{rrr}
\tablecaption{The {\it HST} Observations \label{tabobs}}
\tablewidth{0pt}
\tablehead{
\colhead{Filter} & \colhead{Exposure time (s)} & \colhead{Image name}
}
\startdata
F336W & 400\xxxx\xxx & U54G0107 \nl
F336W & 400\xxxx\xxx & U54G0108 \nl
F555W &  50\xxxx\xxx & U54G0103 \nl
F555W & 140\xxxx\xxx & U54G0104 \nl
F555W & 140\xxxx\xxx & U54G0105 \nl
F555W & 300\xxxx\xxx & U54G010C \nl
F555W & 300\xxxx\xxx & U54G010D \nl
F814W &  50\xxxx\xxx & U54G0106 \nl
F814W & 100\xxxx\xxx & U54G0109 \nl
F814W & 100\xxxx\xxx & U54G010a \nl
F814W & 300\xxxx\xxx & U54G010b \nl
F656N & 800\xxxx\xxx & U54G0101 \nl
F656N & 800\xxxx\xxx & U54G0102 \nl
\enddata
\end{deluxetable}

\clearpage

\begin{table}
\dummytable\label{tabinteg}
\end{table}

\setcounter{page}{22}

\clearpage

\begin{deluxetable}{lccccrcc}
\tablecaption{Cluster Parameters \label{tabking}}
\tablewidth{0pt}
\tablehead{
\colhead{} & \colhead{$\mu_0$} & \colhead{r$_t$}
& \colhead{r$_c$} & \colhead{r$_c$} 
& \colhead{} 
& \colhead{\protect\rhalf\tablenotemark{b}} 
& \colhead{\protect\rhalf} \nl
\colhead{Cluster} & \colhead{(mag)} & \colhead{(\arcsec)} 
& \colhead{(\arcsec)} & \colhead{(pc)} 
& \colhead{c\tablenotemark{a}} 
& \colhead{(\arcsec)} & \colhead{(pc)}
}
\startdata
A  & \nodata & \nodata & \nodata & \nodata & \nodata & 0.19 & 2.3 \nl
B  & 12.48$\pm$0.12 & 4.78$\pm$2.0  & 0.048$\pm$0.009 & 0.58$\pm$0.11 & 2.0$\pm$0.3 & 0.26 & 3.1 \nl
30 & 13.5$\pm$0.5   & 1.25$\pm$0.75 & 0.026$\pm$0.005 & 0.32$\pm$0.06 & 1.7$\pm$0.3 & 0.20 & 2.5 \nl
35 & 15.5$\pm$0.25  & 4.0$\pm$2.0   & 0.062$\pm$0.006 & 0.75$\pm$0.07 & 1.8$\pm$0.2 & 0.18 & 2.2 \nl
Globular cluster\tablenotemark{c} & \nodata & \nodata & \nodata & 2.6$\pm$3.9 & 1.4$\pm$0.4 & \nodata & 3 \nl
\enddata
\tablenotetext{a}{Concentration index log(r$_t$/r$_c$).}
\tablenotetext{b}{Half-light radius in F555W.}
\tablenotetext{c}{Average parameters of Milky Way globular clusters as
seen today. From 
Djorgovski (1993); Pryor \& Meylan (1993); and Trager, Djorgovski, \& King (1993).}
\end{deluxetable}

\clearpage

\begin{deluxetable}{lccc}
\tablecaption{Stellar CO Equivalent Widths \label{tabco}}
\tablewidth{0pt}
\tablehead{
\colhead{} & \colhead{} & \colhead{$^{12}$CO(2,0)}
& \colhead{$^{12}$CO(3,1)} \nl
\colhead{Object} & \colhead{Spectral Type} & \colhead{(\AA)} & \colhead{(\AA)} 
}
\startdata
Cluster A   &       & 25.4 & 19.7 \nl
Cluster B   &       & 26.4 & 18.8 \nl
HR8837      & A0V   &  0   &  0   \nl
HR1040      & A0Ia  &  0   &  0   \nl
$\sigma$And   & A2Va  &  0   &  0   \nl
64 Psc      & F7V   &  0   &  2.3 \nl
HR8752      & G5Ia  &  0   &  0   \nl 
$\xi$And    & KOIII &  9.3 &  7.6 \nl
35$\sigma$Per & K3III & 21.2 & 18.0 \nl
HR8726      & K5Ib  & 33.7 & 22.3 \nl
$\chi$Peg   & M2III & 26.6 & 17.7 \nl
$\mu$Cep    & M2Ia  & 40.6 & 26.5 \nl
\enddata
\end{deluxetable}

\clearpage

\figcaption{Mosaic of the {\it HST} CCD images of the 140 second F555W
exposure of NGC 1569.
\label{figmos}}

\figcaption{False-color display of the
PC image of the center of NGC 1569 shows
the two super star clusters and the surrounding gas.
The medium exposure F555W is shown as blue, the medium exposure
F814W is shown as green, and the H$\alpha$ image is displayed
in red. The orientation is shown in Figure \protect\ref{figmos},
and the scale is shown in Figure \protect\ref{figclus}(a).
\label{figcolor}}

\figcaption{Star clusters discussed in the text are identified
on the WFPC2 images: a) PC1, b) WF2, and c) WF4.
The circle size is the size of the aperture
used in the integrated photometry of the clusters.
\label{figclus}}

\figcaption{Star clusters are shown in a UVI color-color diagram.
Three clusters with (U$-$V)$_0$$>$0.8 and high uncertainties in
that color are not included in the plot. 
The solid curve in the upper panel is an evolutionary track for a cluster with 
instantaneous star formation and a metallicity of 0.004 and
a Salpeter (1955) stellar initial mass function with an upper
limit of 100 M\protect\solar\ (Leitherer \et\ 1999). The
solid curve in the lower panel 
are their models for a metallicity of 0.008.
Ages 1--9 Myrs in steps of 1 Myrs are marked with x's along these lines;
ages 10, 20, and 30 Myrs are marked with open circles.
The evolutionary tracks end at 1 Gyr.
The arrow in the lower left corner of the upper panel is a reddening line for
a change of 0.2 in E(B$-$V)$_t$. It represents the average of an O6 and
a K5 type spectrum with A$_V$/E(B$-$V)$=$3.1 and a Cardelli \et\
(1989) reddening curve.
Clusters A, B, 30, and 35 are plotted as their names rather than as points;
their uncertainties are comparable to or smaller than the symbols.
\label{figclccd}}

\figcaption{Star clusters shown in UVI color-magnitude diagrams.
The solid curve is an evolutionary track for a cluster with 
instantaneous star formation and a metallicity of 0.004 and
a Salpeter (1955) stellar initial mass function with an upper
limit of 100 M\protect\solar\ (Leitherer \et\ 1999); the dashed
line are their models for a metallicity of 0.008.
Ages 1--9 Myrs in steps of 1 Myrs are marked with x's along these lines;
ages 10, 20, and 30 Myrs are marked with open circles.
The evolutionary tracks end at 1 Gyr.
The evolutionary tracks have been scaled to the expected mass of
cluster A; for other masses the lines would slide up and
down in the diagrams. 
Clusters A, B, 30, and 35 are plotted as their names rather than as points;
their uncertainties are comparable to or smaller than the symbols.
\label{figclcmd}}

\figcaption{Contour plots of clusters A and B and two comparison
clusters from the 140 second F555W image.
The tics for clusters A and B mark every 10 pixels which
is 0.455\arcsec. 
The tics for clusters 30 and 35 mark every pixel which is
0.0455\arcsec.
\label{figcontour}}

\figcaption{Color ratio image of (a) cluster A and (b) cluster B.
We have divided the F555W images by the F814W images,
and averaged the short, medium, and long exposure images. Saturated
pixels in the longer exposures have been replaced with values from
the shorter exposures. 
Two contours from the medium exposure F555W image are shown superposed
in order to outline the cluster and delineate the luminosity peaks.
\label{figratio}}

\figcaption{Integrated and surface brightness profiles in
F555W and integrated and annuli color profiles of F555W$-$F814W
and F336W$-$F555W: a) cluster A,
b) cluster B, c) cluster 30, and d) cluster 35.
The solid curves in the surface brightness plots are
the best fit King model.
To facilitate comparison between the clusters, 
for each cluster we show the same range in radius,
a 0.3 magnitude range in each of the integrated colors,
and a 0.5 magnitude range in annuli colors.
Uncertainties are shown for all measurements, but in many
cases the error bars are smaller than the point size.
Magnitudes are not corrected for reddening or red leak.
\label{figprofile}}

\figcaption{Near-infrared spectra of clusters A and B. The
$^{12}$CO(2,0) at 22935 \AA\ and $^{12}$CO(3,1) at 23227 \AA\
stellar absorption features
are prominent.
\label{figco}}

\figcaption{Color-magnitude diagram of resolved stars within
a 30 pixel, 1.36\arcsec, radius of the cluster center:
a) Cluster A and b) Cluster B. The cluster stars in or near
the clusters are shown as large starred points. The small
x's are the entire resolved stellar population that we measured
in NGC 1569. The absolute magnitude shown along the right vertical
axis is for a distance of 2.5 Mpc and an E(B$-$V)$_t$ of 0.56
and the average responses of O6 and K5 spectral energy distributions.
\label{figcmdcl}}

\figcaption{Colors and distance from the cluster center
of resolved stars within a 30 pixel,
1.36\arcsec, radius of the cluster centers for
the super star clusters a) A and b) B.
Colors are not corrected for reddening.
\label{figclradius}}

\end{document}